\documentclass[prl,twocolumn,superscriptaddress,floatfix,preprintnumbers,amssymb,amsmath]{revtex4}
\usepackage{graphicx}% Include figure files
\usepackage{dcolumn}% Align table columns on decimal point
\usepackage{bm}% bold math
\usepackage[latin1]{inputenc}
\usepackage[mathscr]{eucal}
\usepackage{epsfig}
\usepackage{rotating}

\begin{document}
\preprint{ }

\title{Single-mode heat conduction by photons}
%\title{Radiation of heat through a superconducting line}

\author{Matthias Meschke}
\affiliation{Low Temperature Laboratory, Helsinki University of
Technology, P.O. Box 3500, 02015 TKK, Finland}
\author{Wiebke Guichard}
\affiliation{Low Temperature Laboratory, Helsinki University of
Technology, P.O. Box 3500, 02015 TKK, Finland}
\affiliation{University Joseph Fourier and CNRS, B.P. 166, 25 Avenue
des Martyrs, 38042 Grenoble-cedex 09, France}
\author{Jukka P. Pekola}
\affiliation{Low Temperature Laboratory, Helsinki University of
Technology, P.O. Box 3500, 02015 TKK, Finland}

\pacs{}

\maketitle
%\section{Summary paragraph}
{\bf {\sl Electrical} conductance is quantized in units of
$\sigma_{\rm Q}=2e^2/h$ in ballistic one-dimensional conductors
\cite{wees88,wharam88}. Similarly, {\sl thermal} conductance at
temperature $T$ is expected to be limited by the quantum of thermal
conductance of one mode, $G_{\rm Q} = \frac{\pi k_{\rm
B}^2}{6\hbar}T$, when physical dimensions are small in comparison to
characteristic wavelength of the carriers \cite{pendry83}. The
relation between $\sigma_{\rm Q}$ and $G_{\rm Q}$ obeys the
Wiedemann-Franz law \cite{ashcroft76} for ballistic electrons (apart
from factor 2 in $\sigma_{\rm Q}$ due to spin degeneracy)
\cite{greiner97}, but somewhat amazingly the same expression of
$G_{\rm Q}$ is expected to hold also for phonons and photons, or any
other particles with arbitrary exclusion statistics
\cite{rego99,blencowe00}. The single-mode heat conductance is
particularly relevant in nano-structures, e.g., when studying heat
conduction by phonons in dielectric materials \cite{schwab00}, or
cooling of electrons in metals at very low temperatures
\cite{schmidt04}. Here we show, based on our experimental results,
that at low temperatures heat is transferred by photon radiation, in
our case along a superconducting line, when electron-phonon
\cite{roukes85} as well as normal electronic heat conduction are
frozen out. Thermal conductance is limited by $G_{\rm Q}$,
approaching this value towards low temperatures. Our observation has
implications on, e.g., performance and design of ultra-sensitive
bolometers and electronic micro-refrigerators \cite{giazotto06},
whose operation is largely dependent on weak thermal coupling
between the device and its environment.}

\begin{figure}
    \begin{center}
    \includegraphics[width=8.5cm]{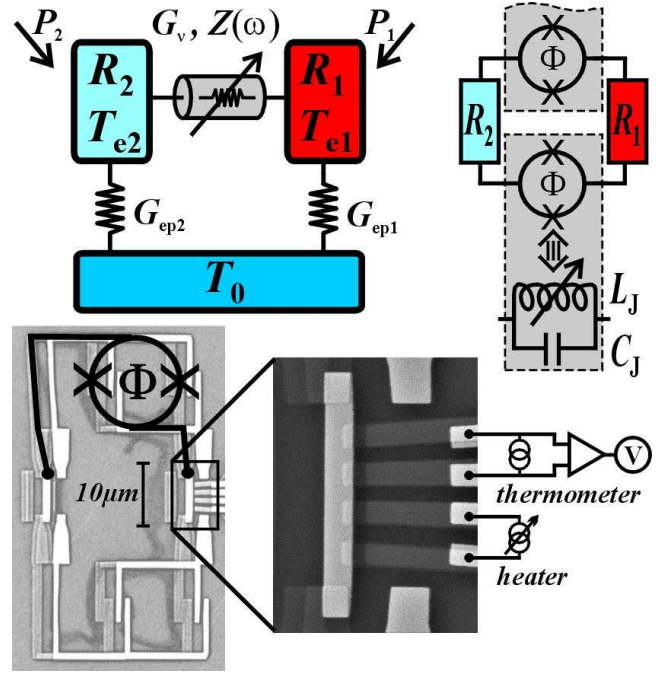}
    \end{center}
    \caption{The system under investigation. On top we show thermal (left)
    and electrical (right)
    models, and on bottom a scanning electron micrograph of the device
    (left), and of resistor 1 with four adjoining NIS and
    two NS contacts
    zoomed (right).}
    \label{fig:Fig1_SEM_thermal_elec}
\end{figure}

To get a picture of the radiative thermal coupling, we start by
considering two resistors at temperatures $T_{{\rm e}1}$ and
$T_{{\rm e}2}$, whose resistances are $R_1$ and $R_2$, respectively,
connected via a frequency ($\omega/2\pi$) dependent impedance
$Z(\omega)$, see Fig. \ref{fig:Fig1_SEM_thermal_elec}. For
simplicity we assume $Z(\omega)$ to be fully reactive, so that only
the two resistors emit and absorb noise heating. The net power flow
$P_\nu$ between the two resistors from 1 to 2 due to the
electron-photon coupling is then given by \cite{pendry83,schmidt04}
\begin{equation} \label{flux}
P_\nu = \int_0^\infty \frac{d\omega}{2\pi} \frac{4R_1R_2}{|Z_{\rm
t}(\omega)|^2}\hbar \omega [n_1(\omega)-n_2(\omega)].
\end{equation}
Here, $Z_{\rm t}(\omega) \equiv R_1+R_2+Z(\omega)$ is the total
series impedance of the circuit, and $n_i(\omega)\equiv [\exp(\hbar
\omega/k_{\rm B}T_{{\rm e}i})-1]^{-1}$ are the boson occupation
factors at the temperatures of the resistors $i=1,2$. Specifically,
for a lossless direct connection of the two resistors,
$Z(\omega)\equiv 0$, we can integrate \eqref{flux} easily with the
result
\begin{equation} \label{flux0}
P_\nu^{Z=0} = r_0 \frac{\pi k_{\rm B}^2}{12\hbar}(T_{{\rm
e}1}^2-T_{{\rm e}2}^2).
\end{equation}
Here $r_0\equiv 4R_1R_2/(R_1+R_2)^2$ is the matching factor, which
obtains its maximum value of unity, when $R_1 = R_2$. Thermal
conductance by the photonic coupling, $G_{\nu}$, defined as the
linear response of $P_\nu$ for small temperature difference $\Delta
T \equiv T_{{\rm e}1}-T_{{\rm e}2}$ around $T\equiv (T_{{\rm
e}1}+T_{{\rm e}2})/2$ can then be obtained from \eqref{flux0} for
the lossless connection as
\begin{equation} \label{cond}
G_{\nu} = r_0 G_{\rm Q}.
\end{equation}
Thus it attains the maximum value for a single transmission channel,
the quantum of thermal conductance, in a matched circuit. This
result is predicted to hold not only for such photon-mediated
coupling, but much more generally for carriers of arbitrary
exclusion statistics \cite{wilczek82,haldane91} from bosons to
fermions \cite{rego99,kane96,kane97}.

Due to its relatively weak temperature dependence, $\propto T$,
electron-photon coupling is expected to become the dominant
relaxation means at sufficiently low $T$. The competing
electron-phonon thermal conductance $G_{\rm ep}$ behaves normally as
$G_{\rm ep} \simeq 5\Sigma \Omega T^4$, where $\Sigma$ is a material
parameter and $\Omega$ is the volume of the resistor. This result
derives from the expression of heat flux from electrons to lattice,
$P_{\rm ep} \simeq \Sigma \Omega (T_{{\rm e}i}^5-T_0^5)$, where
$T_{{\rm e}i}$ and $T_0$ are the temperatures of the electrons in
the resistor and of the lattice, respectively \cite{roukes85}.
Equating $G_\nu$ from \eqref{cond} and $G_{\rm ep}$, one finds the
cross-over temperature, $T_{\rm cr}$, below which the photonic
conductance should dominate: $T_{\rm cr} = [r_0\pi k_{\rm
B}^2/(30\hbar \Sigma \Omega)]^{1/3}$ \cite{schmidt04}. For typical
metals, for which $\Sigma \sim 10^9$ Wm$^{-3}$K$^{-5}$, for
mesoscopic resistors with $\Omega \sim 10^{-20}$ m$^3$, and for
matching where $r_0$ is not too low as compared to unity, one
obtains $T_{\rm cr}\sim 100 - 200$ mK. Such temperatures are in the
range of experiments that we describe here. By state-of-the-art
electron-beam lithography one can obtain metallic islands of volumes
as small as $< 10^{-24}$ m$^3$ \cite{pashkin00}, and there $T_{\rm
cr}$ could be as high as several K.

To investigate the electron-photon thermal conduction experimentally
we have imbedded a tunable impedance between the two resistors. This
allows us to measure the modulation of $P_\nu$, or $G_{\nu}$, in
response to the externally controllable impedance $Z(\omega)$. In
practice the two resistors are connected to each other symmetrically
by two aluminium superconducting lines, interrupted by a DC-SQUID
(Superconducting Quantum Interference Device) in each line, as shown
in the electron micrograph and in the electrical model in Fig.
\ref{fig:Fig1_SEM_thermal_elec}. These SQUIDs serve as the thermal
switches between the resistors as will be described below.

The structures have been fabricated by electron-beam lithography and
three angle shadow evaporation. The film thickness of the
superconducting lines is 20 nm. The two AuPd resistors are nominally
identical, 6.6 $\mu$m long, 0.8 $\mu$m wide and 15 nm thick. Their
resistances are $R_i \simeq 200$ $\Omega$ each. One of them, say
$R_1$, is connected by four NIS (normal-insulator-superconductor)
tunnel junctions to external aluminium leads to allow for
thermometry \cite{nahum93} and Joule heating. The normal state
resistance of each NIS junction is about 50 k$\Omega$. The resistors
are connected by direct NS contacts to the superconducting lines in
between, without a tunnel barrier. The resistors are, however, long
enough such that they are not affected by proximity
superconductivity noticeably. This is verified by the measured
tunnel characteristics of the nearby NIS tunnel junctions. Due to
the superconductors at a low working temperature, which is typically
a factor of ten below the critical temperature $T_{\rm C} \simeq
1.2$ K of aluminium, the normal electronic thermal conductance along
the lines is efficiently suppressed.

Each DC-SQUID can be modelled as a parallel connection of a
Josephson inductance $L_{\rm J}$ and capacitance $C_{\rm J}$.
$L_{\rm J}\simeq \hbar/(2eI_{\rm C})$ can be tuned by external
magnetic flux $\Phi$ threading through the DC-SQUID loop, since the
critical current is $I_{\rm C} \simeq I_{\rm C0}|\cos(\pi
\Phi/\Phi_0)|$. Here $I_{\rm C0}$ is the Ambegaokar-Baratoff
critical current \cite{ambegaokar63} determined by geometry and
materials, and $\Phi_0 = h/2e \simeq 2\cdot10^{15}$ Wb is the flux
quantum. Capacitance $C_{\rm J}$ is a constant determined by
geometry and materials.

The length of the lines connecting resistors $R_1$ and $R_2$ is
$\sim 30$ $\mu$m, i.e., much shorter than the typical thermal
wavelength $\lambda_{\rm th} =2\pi\hbar c/k_{\rm B}T$, which is
several centimetres at 100 mK; here $c\sim 10^8$ m/s is the speed of
light on silicon substrate. Therefore we do not need to consider a
distributed electrical model with a transmission line, but instead
$Z(\omega)$ is effectively a lumped series connection of two $LC$
circuits, i.e., $Z(\omega) = i2\omega L_{\rm
J}/[1-(\omega/\omega_0)^2]$. Here $\omega_0 =(L_{\rm J}C_{\rm
J})^{-1/2}$ and we have assumed for simplicity that the two
DC-SQUIDs are identical and they are exposed to the same magnetic
field. This is expected to be a good approximation in view of the
symmetric experimental configuration. The net heat flow between 1
and 2 is then given by
\begin{equation} \label{fluxLC}
P_\nu =  \frac{\pi k_{\rm B}^2}{12\hbar}(r_1T_{{\rm
e}1}^2-r_2T_{{\rm e}2}^2),
\end{equation}
where the matching parameters $r_i$ now depend on temperature as
\begin{equation} \label{PLC}
r_i = \frac{6r_0}{\pi^2} \int_0^\infty dx \frac {x}{e^x-1}
\big{[}1+\frac{(\omega_{{\rm th},i}\tau_{\rm R})^2
x^2}{[1-(\omega_{{\rm th},i}/\omega_0)^2x^2]^2}\big{]}^{-1}.
\end{equation}
Above we have defined $\omega_{{\rm th},i}\equiv k_{\rm B}T_{{\rm
e}i}/\hbar$ and $\tau_{\rm R} \equiv L_{\rm J}/[(R_1+R_2)/2]$.

For the full description, we still need a thermal model
incorporating the two competing relaxation mechanisms, due photons
and phonons, respectively. This is schematically depicted in Fig.
\ref{fig:Fig1_SEM_thermal_elec}, where the temperature in each
resistor tends to relax via electron-phonon coupling $G_{{\rm ep}i}$
to the constant temperature $T_0$ of the bath, and towards a common
temperature of the two resistors via the tunable photonic
conductance $G_\nu$. $P_i$ denotes the external heat leak into
resistor $i$: due to wire connections $P_1$ has a significant
non-zero value even in the absence of intentional heating, whereas
$P_2$ turns out to be very small, since the corresponding resistor
is not connected directly to external leads. We may describe the
steady state of the system by thermal master equations
\begin{equation} \label{diff1}
P_i=\pm \frac{\pi k_{\rm B}^2}{12\hbar}(r_1T_{{\rm e}1}^2-r_2T_{{\rm
e}2}^2)+\Sigma\Omega_i(T_{{\rm e}i}^5-T_0^5), i=1,2
\end{equation}
where $\pm$ equals $+$ for $i=1$, and $-$ for $i=2$, and $\Omega_i$
is the volume of resistor $i$. Equations \eqref{diff1} combined with
\eqref{PLC} can be solved numerically to obtain the temperatures
$T_{{\rm e}i}$ under given conditions.

The experiments were performed in a $^3$He-$^4$He dilution
refrigerator at temperatures from 30 mK up to several hundred mK.
All the measurement wiring was carefully filtered and essentially
only DC signals were employed. A typical measurement was done as
follows. One of the SINIS tunnel junction pairs connected to
resistor 1 was used as a thermometer by applying a small ($\sim 3$
pA) DC current through it and by measuring the corresponding
temperature dependent voltage. When biased at a low enough current,
high sensitivity is obtained without significant self-heating or
self-cooling effects \cite{giazotto06}. The SINIS thermometer is
then calibrated by measuring this voltage against the bath
temperature $T_0$, by slowly sweeping the temperature of the mixing
chamber through the range of interest, from about 40 mK up to about
300 mK. The actual measurements of thermal coupling are then done by
stabilizing the bath at a desired temperature. Then we apply slow
sweeps ($\sim 1\Phi_0$/min) of external magnetic flux, which threads
the two DC-SQUID loops nominally identically. A typical amplitude of
the field sweep is such that it corresponds to 4 - 5 flux quanta
through each DC-SQUID. In our geometry this corresponds to a field
of about 100 $\mu$T. SINIS thermometer reading is then averaged over
several tens of such field sweeps, to measure accurately the
periodic variation of $T_{{\rm e}1}$ in response to the field sweep.

\begin{figure}
    \begin{center}
    \includegraphics[width=8.5cm]{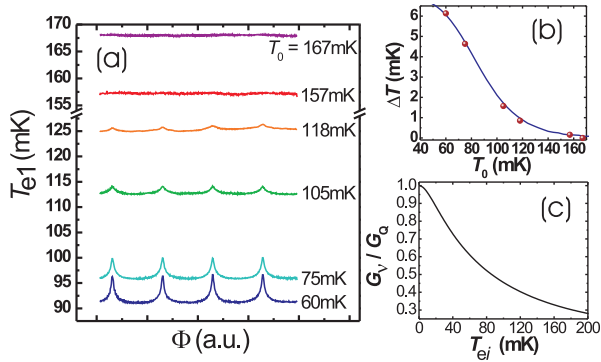}
    \end{center}
    \caption{Flux modulation of the temperature of resistor 1 in the
    absence of external heating. In (a), $T_{{\rm e}1}$ has
    been recorded at several bath temperatures $T_0$ indicated.
    The measured amplitude (symbols)
    of temperature modulation
    is plotted against $T_0$ in (b) and compared to the theoretical model (line) as
    described in the text.
    %At the bottom right, two periods of flux modulation at the
    %base temperature have been zoomed and compared to the theoretical model.
    In (c), the ratio of the photonic thermal conductance $G_\nu$ and
    the quantum of thermal conductance $G_{\rm Q}$ is shown as a function
    of the electron temperature $T_{{\rm e}i}$ using the parameters of our circuit.}
    \label{fig:noheat}
\end{figure}

Figure \ref{fig:noheat} shows results of a measurement at a few bath
temperatures as described above. We see that the modulation of
$T_{{\rm e}1}$ in response to magnetic flux $\Phi$ is about 6 mK at
the lowest bath temperature and it decreases monotonically when
$T_0$ is increased. Based on this data and our electrical and
thermal models above, we expect that the maxima in $T_{{\rm e}1}$
correspond to the weakest electron-photon coupling at half-integer
values of $\Phi/\Phi_0$. In Fig. \ref{fig:noheat} (b) we plot with
circles the corresponding modulation amplitude of $T_{{\rm e}1}$
between its maximum and minimum values, $\Delta T \equiv T_{{\rm
e}1,{\rm max}}-T_{{\rm e}1,{\rm min}}$, as a function of $T_0$ for
the experimental data in (a). We have added to the figure a solid
line from our theoretical model assuming $R_1=R_2$, $P_1 = 1$ fW,
$P_2=0$, $I_{{\rm C}0} = 20$ nA. The last value is a fit parameter,
since we cannot measure it directly in our geometry, but it is in
line with the measured values of critical currents of DC-SQUIDs that
we fabricated with the same parameters separately. We set $C_{\rm J}
= 15$ fF based on the geometry of the device. Additionally, we use a
typical value $\Sigma = 2\cdot 10^9$ WK$^{-5}$m$^{-3}$
\cite{giazotto06}, and $\Omega_i = 6\cdot10^{-20}$ m$^3$ for both
resistors. The latter corresponds to the volume of each resistor
excluding the overlap areas of NS contacts. The agreement between
the experiment and the model is very good and we therefore use these
very realistic parameters in analyzing all the results in this
article. These data imply, see Fig. \ref{fig:noheat} (c), that $r_i$
at integer values of $\Phi/\Phi_0$ lies in the range 0.6...0.3, when
we vary temperature from $\simeq 60$ mK (the approximate value of
$T_{{\rm e}2}$ at the minimum bath temperature) up to 200 mK, i.e.,
the electron-photon conductance is about one half of its quantum
value in our experiment. Note that $G_\nu$ approaches the quantum of
thermal conductance upon lowering temperature, since the thermal
frequency $\omega_{{\rm th},i}$ decreases linearly with $T_{{\rm
e}i}$, and the line impedance at low frequencies, determined by
$L_{\rm J}$, decreases likewise.
\begin{figure}
    \begin{center}
    \includegraphics[width=8.5cm]{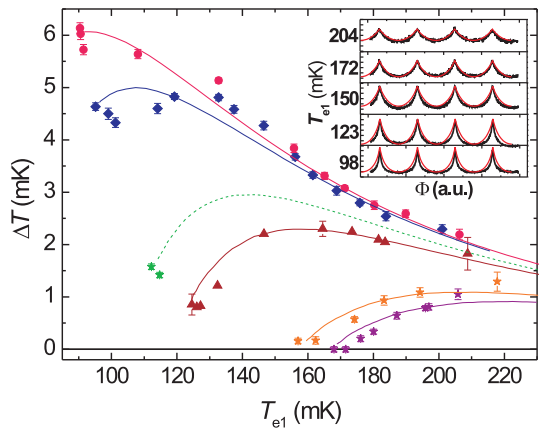}
    \end{center}
    \caption{Results of measurement of $\Delta T$, the
    modulation of $T_{\rm e1}$ when flux is varied,
    with variable amounts of Joule heating applied to resistor 1.
    Different sets of data and lines correspond to different bath
    temperatures $T_0$. From top to bottom
    $T_0 =$ 60 mK, 75 mK, 105 mK, 118 mK, 157 mK and 167
    mK. The symbols refer to experimental data, and lines are from
    the theoretical model described. The inset shows primary data in form of
    $T_{\rm e1}$ against flux, under different input power levels and at $T_0 = 75$ mK.
    Black lines are from the experiment and red ones from the theory.
    The full temperature range in each panel is 6 mK.}
    \label{fig:varT0}
\end{figure}

Next we present experiments where external power was applied to
resistor 1, using the second pair of tunnel junctions connected to
it as a heater. Figure \ref{fig:varT0} demonstrates such
measurements at different bath temperatures. Compared to the data at
the lowest bath temperature of 60 mK where essentially monotonic
decrease of $\Delta T$ can be observed, the intermediate bath
temperature data demonstrates non-monotonic dependence with initial
increase of the signal on increasing the heating, and then slow
decrease towards higher power levels and $T_{\rm e1}$. Finally, at
the highest values of $T_0$, the signal is first absent but emerges
upon increasing the input power. This behaviour arises because by
applying Joule heating to just one resistor, we can establish a
larger temperature difference between the two. Yet at large enough
levels of $P_1$, the overall temperature of the system increases,
and the significance of the photon coupling with respect to
electron-phonon coupling is diminishing, and the temperature
modulation becomes very weak. All these dependences are fully
consistent with our theoretical modelling of the system, and we
obtain quantitative agreement with the data by using the same
electrical and thermal parameters of the system as when modelling
data of Fig. \ref{fig:noheat}. The theoretical lines in Fig.
\ref{fig:varT0} are the result of this modelling. In the inset of
Fig. \ref{fig:varT0} we show a few full modulation curves of $T_{\rm
e1}$ vs. $\Phi$ at the bath temperature of 75 mK. The black lines
are from experiment and the red ones from the theoretical model. The
theoretical lines catch the main features of the shape of the
experimental curves.
%One possible
%reason for the deviation could be that the basic expression of
%$I_{\rm C}$ vs. $\Phi$ given earlier might not hold perfectly.
%Furthermore, we have ignored the influence of the possibly non-zero
%impedance of the superconducting lines and NS contacts.

Prior to our experiment, the electron system like the one discussed
in this article was assumed to be efficiently decoupled from the
environment once the electron-phonon coupling is suppressed at low
temperatures. Therefore, the observed $\propto T$ photonic
conductance has implications on performance and design of
micro-bolometers and calorimeters, where efficient suppression of
thermal coupling is usually taken for granted. Due to better
coupling to the environment by photon conductance, the expected
sensitivity of such devices is reduced, and their noise is enhanced.
On the other hand, this mechanism could possibly provide a way to
tune the thermal coupling of a bolometer to the heat bath in order
to optimize its operation, which is a trade-off between sensitivity
and bandwidth. The radiation of heat can possibly be benefitted also
in removing excessive heat, e.g., on dissipative shunt resistors of
ultra-sensitive or very low temperature SQUIDs \cite{savin06}, or at
the back side of an electronic micro-refrigerator \cite{clark04}.
Furthermore, the photonic coupling could act as a mediator of
decoherence, e.g., on solid-state quantum coherent devices
\cite{schon01}. The strength of this harmful effect depends, like in
our present experiment, critically on matching between the noise
source and the system that it affects on.

{\bf Acknowledgements} We thank Mikko Paalanen, Dmitri Averin, Arttu
Luukanen, Hugues Pothier, Frank Hekking and Gerd Sch\"on for useful
comments, and Academy of Finland (TULE) and the EC-funded ULTI
Project, Transnational Access in Programme FP6 (Contract
\#RITA-CT-2003-505313) for financial support.

\end{document}